\documentclass{article}

\usepackage{lineno,hyperref}
\usepackage{amssymb}
\usepackage{multirow}
\usepackage{caption}
\usepackage{subfigure}
\usepackage{graphicx}
\usepackage{url}
\usepackage{authblk}
\usepackage{setspace}

\title{Predicting Learning Status in MOOCs using LSTM}
\author[a]{Zhemin Liu}
\author[a]{Feng Xiong}
\author[a]{Kaifa Zou}
\author[a]{Hongzhi Wang \thanks{Corresponding author: wangzh@hit.edu.cn}}
\affil[a]{School of computer science and technology Harbin Institute of Technology, Harbin, China}

\begin{document}
\begin{spacing}{1.0}
\maketitle

\begin{abstract}
	Real-time and open online course resources of MOOCs have attracted a large number of learners in recent years. However, many new questions were emerging about the high dropout rate of learners. For MOOCs platform, predicting the learning status of MOOCs learners in real time with high accuracy is the crucial task, and it also help improve the quality of MOOCs teaching. The prediction task in this paper is inherently a time series prediction problem, and can be treated as time series classification problem, hence this paper proposed a prediction model based on RNN-LSTMs and optimization techniques which can be used to predict learners' learning status. Using datasets provided by Chinese University MOOCs as the inputs of model, the average accuracy of model's outputs was about 90\%.\\
	keywords:MOOCs, Behavior Prediction, LSTMs
\end{abstract}

\section{INTRODUCTION}
Massive open online courses (MOOCs) have the potential to enable free education on an enormous scale, since the course material, video lectures, handouts and reading materials are accessible to anyone who wants to study \cite{kay2013moocs}. In addition, MOOCs' learners have the possibility to engage in discussions or ask questions in associated forums, leading to the creation of a community among learners and professors, which reflects the interactive and engaging character of MOOCs. Hence, MOOCs will be a revolution in delivering high quality learning opportunities to the public.
\par
The participants of MOOCs is enormous, but a major concern often raised about MOOCs is that the completion rate of courses is very low \cite{khalil2014moocs}. According to statistics, the average completion rate of Chinese University MOOCs courses was only 1.5\% \cite{beijingNews}, which mostly means vast majority of learners who signed up the course in the beginning cannot complete the course learning. It would be valuable for professors to know how likely a learner would drop out during the course, they can make some adjustments in time during the teaching process, e.g. send some email reminders or give some positive feedback to learners who are believed to be very likely to drop out during the course. Hence, it is a key task to accurately predict whether a learner is likely to drop out in the near future for promoting the interest of learners in courses.
\par
The prediction of learners learning status brings a number of difficulties.
\par
Firstly, the learners hold various learning purposes, some of whom never learn the course before, and are vulnerable to the difficulty of the curriculum. However, some of them have acquired the knowledge of the curriculum, and choose a few of the courses to take, but even a few of them only take part in the final examination. These factors bring noise into the dataset.
\par
Secondly, the back-end servers of Chinese University MOOCs platform record the learning behaviour data on a large scale, so it is a key point to perform valuable and effective selection among a large amount of data.
\par
Thirdly, since there is a wide gap in the ratio between the learners who give up in the middle of the course and the learners who complete the course according to the dataset provided by Chinese University MOOCs, solving the imbalance problem is also a critical technical point.
\par
The prediction of learners behaviour is inherently a task of time series prediction, which utilizes the learners historical data to predict the learners learning status in the future. A great number of research work has been done on this topic. Related solutions proposed can be divided into statistical analyses based and machine learning based.
\par
Most prior research work applied statistical methods. Logistic regression analysis are applied in \cite{jiang2014predicting}\cite{kizilcec2015attrition}\cite{koedinger2015learning}\cite{taylor2014likely}\cite{whitehill2015beyond}\cite{ye2014early} to predict dropout-prone learners. In addition, Balakrishnan\cite{balakrishnan2013predicting} leveraged Hidden Markov Model(HMM) to solve the prediction problem with features extracted from discussion forums and video lectures.
\par
Besides the statistical methods, some machine learning based methods also are proposed. Kloft \cite{kloft2014predicting} proposed a dropout predicting method based on support vector machine(SVM), but this proposed method solely focuses on click-stream data.
\par
However, the prior related approaches present shortcomings. It's a tough task to determine acceptance/rejection thresholds in logistic regression analysis approaches to interpret student-progress attributes which may have negative impact on the final results. As the experiments demonstrated in \cite{tian2015predicting}, When complexity of SVM is similar to that of recurrent neural network with memory blocks, the results acquired by SVM are far from those of neural network.  The HMM also has some major limitations: it requires both forward pass and backward pass through the entire data sequence, and it is assumed that the next state is dependent only upon the current state which is not a good idea for time series prediction task. In order to achieve higher prediction accuracy, the prediction model in this paper applied neural network, which can tackled the above-mentioned shortcomings.
\par
The recurrent neural networks which differs from traditional neural network is to make use of sequential information. They perform the same task for each element of a sequence, with the output being dependent on the previous computations. Long Short-Term Memories(LSTMs) is a specific RNN architecture designed to model temporal sequences and their long-range dependencies more accurately than conventional RNNs. Hence, we build a prediction model based on RNN-LSTMs, and a RNN-LSTMs for MOOCs allows us to infer a learners status in the next time step based on his/her previous status and their currently observable actions.
\par
This paper makes the following contribution.
\par
Firstly, We transform the dropout prediction problem into a time series prediction problem, and utilize LSTMs to cope with this problem.
\par
Secondly, for feature extraction, we form features from back-end records data which not only contains learners interaction on forums, but also contains learners interaction with resources and assignments.
\par 
The last but not least, we conduct extensive experiments on real data. Experimental results demonstrate the accuracy of our proposed prediction model, the average accuracy of prediction is about 90 percent. Although different definition of dropout acquired slightly various results, the accuracy of prediction begin to rise in most courses since second or third week.
\par 
The rest of the paper is organized as follows. We begin with the formulation of problem, which contains the different assumptions we make in the problem definition. Then, the prediction model based on RNN-LSTMs is presented in section\ref{sec_prediction_model_based_on_LSTMs}, this section describes general workflow, feature extraction and selection, RNN-LSTMs and prediction model specifically. Also the section \ref{sec_prediction_model_based_on_LSTMs} outlines the key technical point and dropout technique in this paper.  Next, we describe the details of the datasets we use in section\ref{sec_datasets_description}, and demonstrate the experimental results of the proposed prediction models with datasets provided by Chinese University MOOCs in next section\ref{sec_experiment}. Finally, we make conclusion of our work in this paper.

\section{FORMULATION OF PROBLEM}
In this section, we formulate the problem of learning status prediction.
\subsection{Timing Step of Prediction}
\label{time_step_of_prediction}
It is the first step to determine the time unit of prediction before making a model to predict learning status of learners. Taking Big Data Algorithm Course in 2014 fall semester as an example (see table \ref{table_publish_time_of_big_data_algorithm}), the professor deliveries lecture videos and corresponding assignments on Mondays or Tuesdays. Because of that, it is reasonable to assume that an active learner would generate at least one studying interactive record. Hence, we regard one week as a time step of prediction and aggregate all student activities within a week for each student.

\begin{table}[htbp]
	\begin{center}
		\caption{Publish time of Big Data Algorithm Course}
		\label{table_publish_time_of_big_data_algorithm}
		\begin{tabular}{l l l}
			\hline
			\textbf{Lecture} & \textbf{Video Release Date} & \textbf{Assignment Release Date} \\
			\hline
			Lecture 1 & 2014-06-30 & 2014-06-30\\
			Lecture 2 & 2014-07-07 & 2014-07-07\\
			Lecture 3 & 2014-07-14 & 2014-07-15\\
			Lecture 4 & 2014-07-22 & 2014-07-23\\
			Lecture 5 & 2014-07-29 & 2014-07-30\\
			Lecture 6 & 2014-08-05 & 2014-08-06\\
			Lecture 7 & 2014-08-12 & 2014-08-13\\
			Lecture 8 & 2014-08-19 & null\\
			Lecture 9 & 2014-08-26 & 2014-08-27\\
			Lecture 10 & 2014-09-02 & 2014-09-03\\
			\hline
		\end{tabular}
	\end{center}
\end{table}

\subsection{Definition of Learning Status}
\label{definition_of_learning_status}
Our primary feature is whether a learner has dropped the course, which is the feature we eventually want to be able to predict and is encoded as a binary value, i.e. a learner can be an active participant or have dropped out.
\par
We consider four definitions of dropout, as summarized in table \ref{table_def_of_learning_status}, to capture different contexts of the learning status in a course. These definitions will be used for later evaluation of prediction model.

\begin{table}[htbp]
	\begin{center}
		\caption{Definition of Learning Status}
		\label{table_def_of_learning_status}
		\begin{tabular}{p{3cm} p{7.7cm}}
			\hline
			\textbf{Definition Num} & \textbf{Detailed Description} \\
			\hline
			DEF1 & Whether a learner has records of learning activities in the comming week? If yes, the status of the learner is active, or dropout.\\
			DEF2 & Whether a learner has records of learning activities in the subsequent weeks? If yes, the status of the learner is active, or dropout.\\
			DEF3 & Whether a student has records of  learning activities in the final week? If yes, the status of the learner is active, or dropout.\\
			DEF4 & Whether a student take the final exam? If yes, the status of the learner is active, or dropout.\\
			\hline
		\end{tabular}
	\end{center}
\end{table}

The DEF1 reflects some continuous status. Based on this definition, corresponding label of a learner in a time step could be generated. And the professor is able to provide instant feedback and interventions. The DEF2, DEF3 and DEF4 are similar in that they both correspond to final status of a learner. Hence the dropout label cannot be defined until the end of the course\cite{he2015identifying}. As for DEF2, a prediction would be made every week for each learner to predict whether the learner will have no learning activities after the current week. For all four learners status definitions, we use label 0 to refer to dropout, and use the label 1 to refer to active.
\par
In addition, the final exams of MOOCs course usually last longer. Hence we take the week before the final exam as the final week in prediction task.

\subsection{Dropout Prediction Formulation}
\label{dropout_prediction_formulation}
We regard the dropout prediction problem as a sequence labelling problem in which both the input features and the output labels are represented as sequences.
\par
As mentioned in \ref{time_step_of_prediction},  we take a week as a time step in our prediction task. Based on this assumption, the input feature of a learner at each time step is generated from all learning activities within a week. According to the definition of learning status presented in \ref{definition_of_learning_status} of this section, we tag the output labels based on the input features of that week. Hence, for a learner, we obtain his/her weekly learning activity statistics in the course, denoted by an input sequence ($x_1$,...,$x_t$), and the corresponding labels forming an output sequence ($y_1$,...,$y_t$), $x_i$ stands for input feature vector of week $i$, $y_i$ stands for the corresponding output label.

\section{Prediction Model based on LSTMs}
\label{sec_prediction_model_based_on_LSTMs}
\subsection{General Workflow}
\begin{figure}[htbp]
	\centering
	\includegraphics[width=4cm]{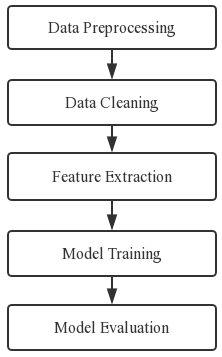}
	\caption{General Workflow}
	\label{figure_general_workflow}
\end{figure}

This paper constructs prediction model of learning status in Chinese University MOOCs with deep learning technique. Modelling process based on deep learning usually contains three phrases: feature selection and extraction, model training, and model evaluation\cite{Liuwenyan2016MOOC}. Datasets from back-end server contain a lot of dirty data, so data preprocessing and data cleaning are necessary steps before the next processing step. After the completion of data preprocessing and data cleaning tasks, we could do feature extraction and selection,  and then build and train the time sequence prediction model based on RNN-LSTMs and the feature sets. Finally, the generated predicting model should be evaluated by the test datasets. The general workflow shown as figure \ref{figure_general_workflow}.

The data from back-end database server are incomplete, noisy and inconsistent, so we utilized data cleaning technique to deal with missing values, smooth noisy data, identify and remove outliers. For repeating and conflict data, we choose to filter them directly. And for incomplete records, we choose to remove them. Also we perform aggregation task by data transformation. In addition, we quantitatively characterize learners online behaviour from web logs and click stream data.
\par
To better gauge learners' intent, we must form variables capturing learners' behaviour. Among many different type of variables, we focus on per-learner longitudinal variables, which involves repeated observation of the same variables over time. A variable is usually an aggregate or a statistic of some observations defined for that time period. In the context of MOOCs, we defined the time interval to be a week which described in section\ref{time_step_of_prediction} in detail. In section\ref{sub_feature_extraction_and_selection}, we present a feature list containing fifteen features, and generate features vectors with statistic technique.
\par
The objective of training model is to generate outputs with training data as inputs which fits well to the pre-defined labels. So in the training process, we take the features vectors from the feature extraction and pre-defined labels as inputs of model. This paper takes the RNN with LSTM cells as hidden units, in the model training, hidden states learned week by week to capture sequential information, which is able to retain past information. This paper also takes the dropout technique to prevent overfitting in training process. As the experiment result shown in section\ref{sec_experiment}, the usage of RNN-LSTMs and dropout technique helps us achieve the training objective.
\par
After the training step, the final step evaluates the performance of the prediction model. We take the test data as the input of the trained model, and obtain corresponding labels for each data point in test data, by comparing the output labels and true labels, the accuracy of the prediction model are computed.

\subsection{Feature Extraction and Selection}
\label{sub_feature_extraction_and_selection}
When a learner takes part in the learning activity of MOOCs platform, various learning status data would be stored in the back-end server, such as lecture video views, class discussion participates and quiz participates, and so on. Extracting effective feature from these recording data is beneficial to the model training. By analysing the learning status of Chinese University MOOCs' learners, fifteen features were extracted:

\begin{table}[htb]
	\begin{center}
		\caption{Feature List}
		\label{table_feature_list}
		\begin{tabular}{l l}
			\hline
			\textbf{ID} & \textbf{Feature(Each Week)} \\
			\hline
			1 & Post Count in General Discussion\\
			2 & Post Count in Professor Answer Area\\
			3 & Post Count in Class Exchange Area\\
			4 & Comments Count in General Discussion\\
			5 & Comments Count in Professor Answer Area\\
			6 & Comments Count in Class Exchange Area\\
			7 & Evaluated Count in Peer-to-Peer Evaluation\\
			8 & Evaluating Count in Peer-to-Peer Evaluation\\
			9 & Views Count of Lecture Content\\
			10 & Views Count of Forum Detail Information\\
			11 & Response to Post Count in General Discussion\\
			12 & Response to Post Count in Professor Answer Area\\
			13 & Response to Post Count in Class Exchange Area\\
			14 & Reply Count in Forum\\
			15 & Quiz Count\\
			\hline
		\end{tabular}
	\end{center}
\end{table}

\subsection{RNN-LSTMs}
\label{subsec_rnn_lstms}
\begin{figure}[htbp]
	\centering
	\includegraphics[scale=0.8]{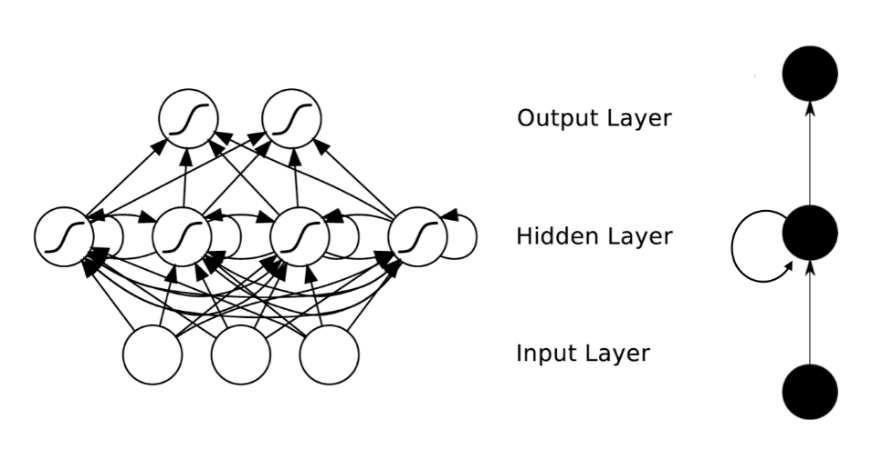}
	\caption{topological structure of RNN\cite{csdnblog}}
	\label{figure_topological_structure_of_RNN}
\end{figure}

RNN is an effective tool to handle temporal sequences. Traditional neural network makes an assumption that the sequences segments inputed into the network are independent, based on this assumption, traditional neural network is unsuitable to modeling on temporal sequences. However, an input in RNN not only depends on current input, but also previous input, which means that the computation of current output depends on historical input sequences and network status information. As shown in figure \ref{figure_topological_structure_of_RNN}, the main idea of RNN is to add loops into the hidden layers \cite{csdnblog}.

\begin{figure}[htbp]
	\centering
	\includegraphics[scale=0.3]{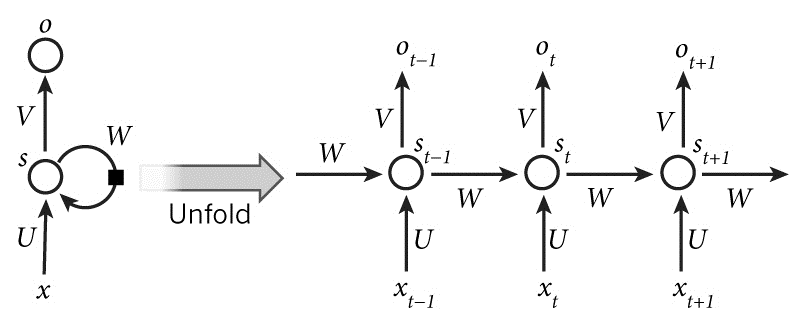}
	\caption{An unrolled RNN\cite{dennyrnn}}
	\label{figure_unrolled_RNN}
\end{figure}

As explained in artical\cite{dennyrnn}, a RNN can be thought of as multiple copies of the same network, each passing a message to a successor, an unrolled RNN described in figure \ref{figure_unrolled_RNN}. It is assumed that the length of input sequence $X$ is $T$, the dimension of input data $X_t$ in a time step is $D$, nodes number of hidden layer is $H$, nodes number of output layer is $K$.

Long Short-Term Memory (LSTM) is a specific RNN architecture that was designed to model temporal sequences and their long-range dependencies more accurately than conventional RNN.
\par
The LSTM contains special units called memory blocks in the recurrent hidden layer, as shown in artical\cite{csdnblogrnnlstm} . The memory blocks contain memory cells with self-connections storing the temporal state of the network in addition to special multiplicative units called gates to control the flow of information. Each memory block in the original architecture contained an input gate and an output gate. The input gate controls the flow of input activations into the memory cell. The output gate controls the output flow of cell activations into the rest of the network. Later, the forget gate was added to the memory block\cite{gers1999learning}. This addressed a weakness of LSTM models preventing them from processing continuous input streams that are not segmented into subsequence. The forget gate scales the internal state of the cell before adding it as input to the cell through the self-recurrent connection of the cell, therefore adaptively forgetting or resetting the cell's memory. In addition, the modern LSTM architecture contains peephole connections from its internal cells to the gates in the same cell to learn precise timing of the outputs\cite{gers2002learning}.

\begin{figure}[htbp]
	\centering
	\includegraphics[width=0.6\linewidth]{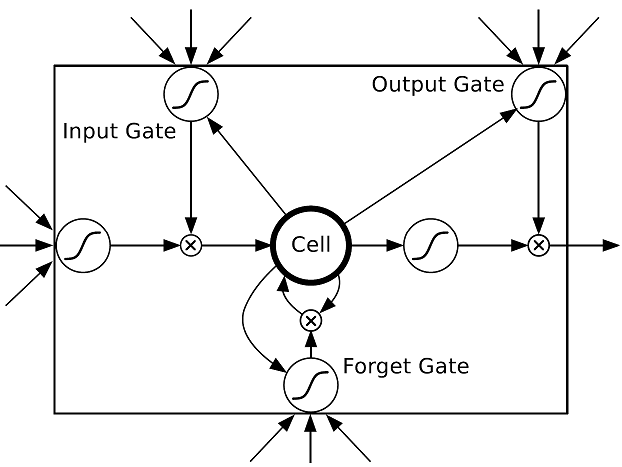}
	\caption{A LSTM CELL\cite{csdnblogrnnlstm}}
	\label{figure_lstm_cell}
\end{figure}

As shown in figure \ref{figure_lstm_cell}, the memory cell in LSTM cell could store historical information, at the same time, input gate, forget gate and output gate are imported to control the update and use of historical information.

\begin{figure}[htbp]
	\centering
	\includegraphics[scale=0.7]{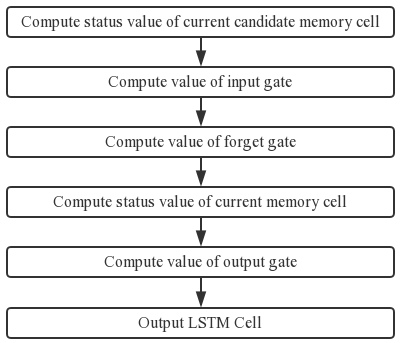}
	\caption{Update process of LSTM Cell}
	\label{figure_update_process_of_LSTM}
\end{figure}

The figure \ref{figure_update_process_of_LSTM} describes the update process of LSTM Cell, which shows the design of three kinds of gate and independent memory cell to implement the storage, reading, resetting and updating of long distant historical information.

\subsection{Learning behaviour prediction based on RNN-LSTM}
\begin{figure}[htbp]
	\centering
	\includegraphics[scale=0.65]{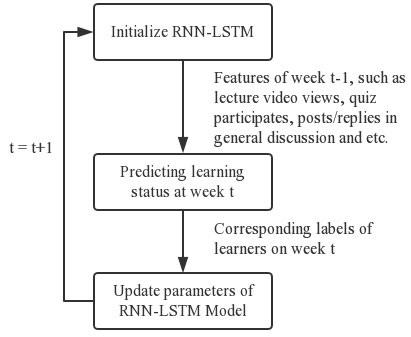}
	\caption{Learning status prediction model based on RNN-LSTM}
	\label{figure_learning_status_prediction_model_based_on_LSTM}
\end{figure}

This paper build prediction model of Chinese University MOOCs learners' learning status based on RNN-LSTM. The model structure is described in figure \ref{figure_learning_status_prediction_model_based_on_LSTM}. Firstly, initialize the prediction model, and then update the parameters of model based on the feature and label sets.

\subsection{Key Technical Point}
\begin{enumerate}
	\item Learning the long memory features in the learners' interactive data by using the memory ability for historical information, and applying these features to the task of predicting the learners' learning status in Chinese University MOOC.
	\item By the utilization of RNN-LSTM's online learning, we update the RNN-LSTM model with latest input data, in the meanwhile, making the prediction model forget the historical information for too long. In this way, the model could reflect current learning status of learners, and the accuracy of the model output would be improved.
	\item Considering the imbalance between the dropout learners and completed learners, during the feature data extraction, this paper transform the value of feature which bigger than 1 to 1.
\end{enumerate}

\subsection{Dropout Technique}
Hinton proposed dropout technique which used to prevent overfitting in neural network in 2012\cite{srivastava2014dropout} dropout in model training will make a certain proportion of hidden nodes unworkable, weights corresponding to those nodes won't update in the training, however the use of the model will utilize all nodes.

\section{Datasets Description}
\label{sec_datasets_description}
Datasets in this paper are provided by Chinese University MOOCs Platform, which contains back-end access records of six courses: A Primer in Game Theory (20001), Finance (21011), Art in china ancient architecture (43002), Sport and Health (45002), First aid general knowledge (85001). Each course varied in course open time, course weeks, and number of registered learners, more detailed information as shown in  Table \ref{table_detailed_information_of_course}. Access data contains user basic information, course selection information, course table, semester table, class hour table, imooc quiz table, imooc exam table, imooc topic table, imooc topic detail table, imooc evaluation table, imooc peer-to-peer evaluation table, imooc peer-to-peer evaluation detail table, logging table.
\par
When a learner participates in the learning activities of MOOCs, his/her all click operations will be recorded in log table by web server, hence log table contains an amount of learner's clickstream data. These data consists of URL, click events, access source and enter page, etc. By making the most of the clickstream data, we can extract learning status data of learners, such as lecture video views, participates in lecture discussion, and participates in quiz.
\par
What's more, by analyzing imooc discussion table, imooc comments table and imooc peer-to-peer evaluation relation table, we can acquire corresponding learning behavior data with the help of statics technique. Also, valuable learning status data could be extracted by combining the log table and data selection technique. When both data are combined, learning behavior datasets which suitable for learning status prediction model could be acquired.

\begin{table}[htbp]
	\begin{center}
		\caption{Detailed Information of Courses}
		\label{table_detailed_information_of_course}
		\begin{tabular}{p{2.7cm}p{1.9cm}p{1.6cm}p{1.8cm}p{1.8cm}p{1cm}}
			\hline
			\textbf{Course} & \textbf{Open Time} & \textbf{Course Weeks} & \textbf{Register Number} & \textbf{Pass Number} & \textbf{Pass Rate} \\
			\hline
			\multirow{2}{2.7cm}{\textbf{Big Data Algorithm}} & 9003 & 13 & 18648 & 10 & 0.054\\
			\cline{2-6} & 253002 & 10 & 36336 & 8 & 0.022 \\
			\hline
			\multirow{4}{2.7cm}{\textbf{A Primer in Game Theory}} & 20001 & 8 & 42918 & 513 & 1.195\\
			\cline{2-6} & 250004 & 10 & 15165 & 552 & 3.640 \\
			\cline{2-6} & 407001 & 12 & 41429 & 717 & 1.731 \\
			\cline{2-6} & 1001620007 & 8 & 36831 & 476 & 1.292 \\
			\hline
			\multirow{2}{2.7cm}{\textbf{Finance}} & 21011 & 10 & 129462 & 771 & 0.596 \\
			\cline{2-6} & 440004 & 10 & 139053 & 488 & 0.351 \\
			\hline
			\multirow{3}{2.7cm}{\textbf{Art in China Ancient Architecture}} & 231002 & 8 & 28019 & 1196 & 4.269 \\
			\cline{2-6} & 457001 & 10 & 29201 & 677 & 2.318 \\
			\cline{2-6} & 50001 & 13 & 9617 & 3060 & 31.82 \\
			\hline
			\multirow{4}{2.7cm}{\textbf{Sport and Health}} & 255006 & 13 & 3675 & 305 & 8.3 \\
			\cline{2-6} & 419003 & 13 & 10342 & 2828 & 27.34 \\
			\cline{2-6} & 1001596003 & 13 & 7614 & 457 & 6.002 \\
			\cline{2-6} & 117001 & 8 & 13364 & 928 & 6.944 \\
			\hline
			\multirow{4}{2.7cm}{\textbf{First aid general knowledge}} & 253007 & 4 & 11369 & 1508 & 13.26 \\
			\cline{2-6} & 357003 & 4 & 35775 & 1727 & 4.827 \\
			\cline{2-6} & 485002 & 4 & 32882 & 880 & 2.676 \\
			\cline{2-6} & 1001584004 & 4 & 21571 & 1017 & 4.715 \\
			\hline
		\end{tabular}
	\end{center}
\end{table}

\section{Experiment}
\label{sec_experiment}
\subsection{Model Parameters Setting}
We introduce Adam Optimizer into the model training process, which is a method for efficient stochastic optimization that only requires first-order gradients with little memory requirement\cite{kinga2015method}. Adam Optimizer is also an algorithm for the optimization of random objective function, random objective function means that objective functions are different at each iteration, the error computation doesn't rely on all records at once, but rely on part of data at each iteration defined as mini-batch\cite{cnblogsadam}, batch size of mini-batch is setted at 100. In addition, the learning rate of optimizer is setted at 0.001, value of dropout is 0.5, number of hidden layer is 2, number of hidden nodes is 200.

\subsection{Results of Model Training}
We built learners' status prediction model based on RNN-LSTM, and implemented the model with TensorFlow provided by Google. Firstly, we used the application interfaces of TensorFlow to define the whole network structure. Secondly, the preprocessed datasets were splitted into two parts, eighty percent of it was used as training datasets, twenty percent of it was used as testing datasets. After the iterative training, the testing datasets were used to evaluate the accuracy of the prediction model.

\begin{figure}[htbp]
	\begin{minipage}[t]{0.5\linewidth}
		\centering
		\includegraphics[width=2.3in,height=1.1in]{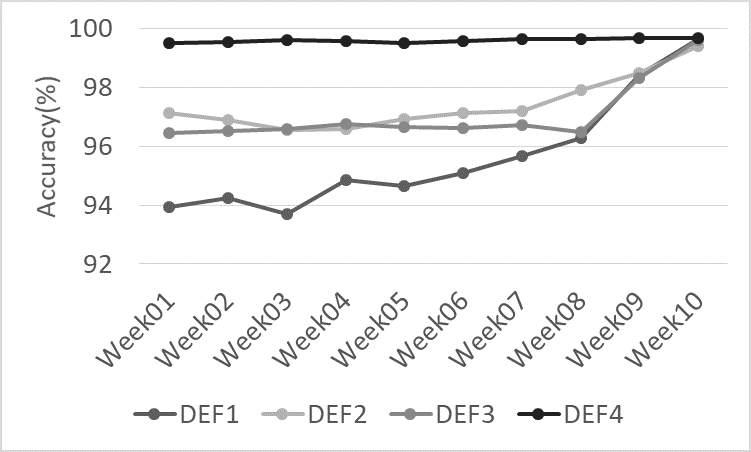}
		\caption{Accuracy of Finance in Term 440004}
		\label{fig:side:a11}
	\end{minipage}
	\begin{minipage}[t]{0.5\linewidth}
		\centering
		\includegraphics[width=2.3in,height=1.1in]{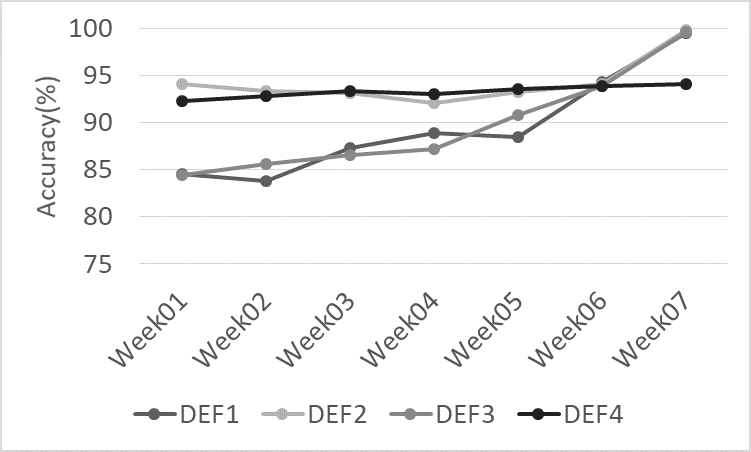}
		\caption{Accuracy of Art in China Ancient Architecture in Term 231002}
		\label{fig:side:b11}
	\end{minipage}
\end{figure}

\begin{figure}[htbp]
	\begin{minipage}[t]{0.5\linewidth}
		\centering
		\includegraphics[width=2.3in,height=1.1in]{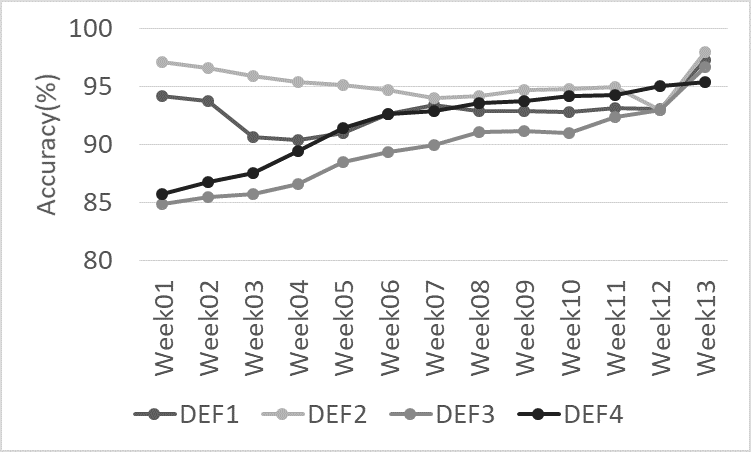}
		\caption{Accuracy of Sport and Health in Term 419003}
		\label{fig:side:a21}
	\end{minipage}
	\begin{minipage}[t]{0.5\linewidth}
		\centering
		\includegraphics[width=2.3in,height=1.1in]{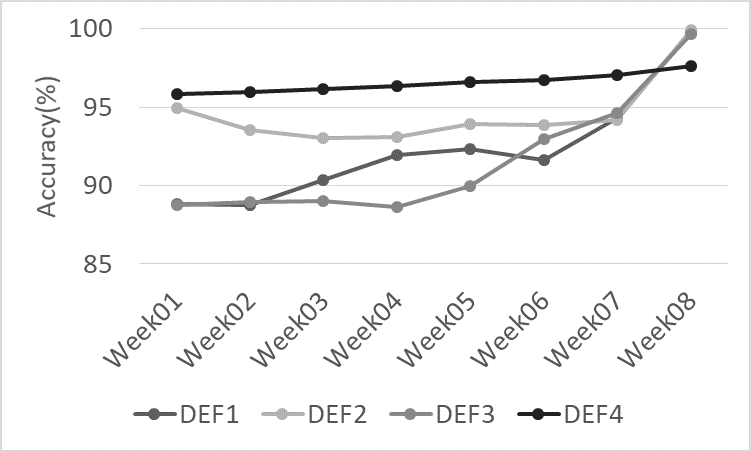}
		\caption{Accuracy of A Primer in Game Theory}
		\label{fig:side:b22}
	\end{minipage}
\end{figure}

\begin{figure}[htbp]
	\begin{minipage}[t]{0.5\linewidth}
		\centering
		\includegraphics[width=2.3in,height=1.1in]{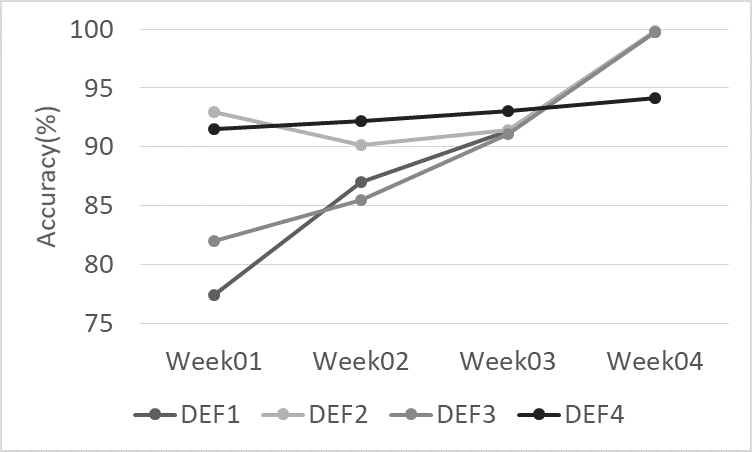}
		\caption{Accuracy of First Aid Common Knowledge in Term 253007}
		\label{fig:side:a31}
	\end{minipage}
	\begin{minipage}[t]{0.5\linewidth}
		\centering
		\includegraphics[width=2.3in,height=1.1in]{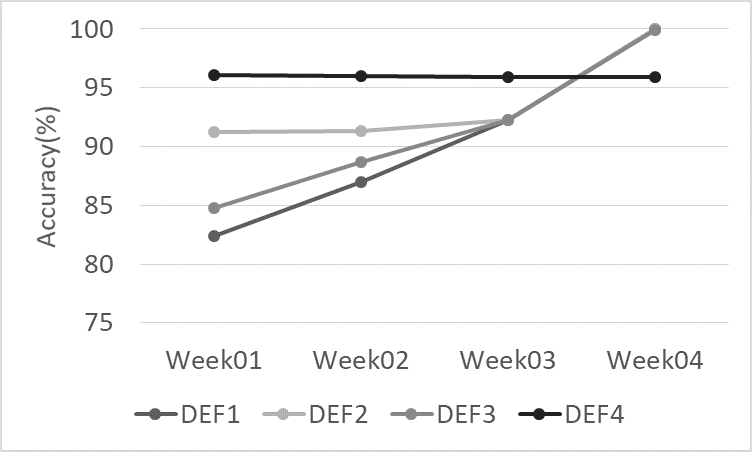}
		\caption{Accuracy of First Aid Common Knowledge in Term 357003}
		\label{fig:side:b32}
	\end{minipage}
\end{figure}

The above figures directly present the predicting results of five courses: First Aid General Knowledge, Art in China Ancient Architecture, Sport and Health, Finance, A Primer in Game Theory. On the whole, accuracy of DEF4 is higher than other three definitions, and relatively more stable. DEF1, DEF2 and DEF3 begin to rise since second week or third week, the average accuracy of each course is 90\%.

\section{Conclusion}
In order to solve the problem of high dropout rate in Chinese University MOOC, it is a primary task to implement the prediction of learners' status. The key task of this paper is to build  prediction model, and get the prediction results of learners' status in this MOOC platform. This prediction problem is abstracted as time series prediction problem, based on this, we build the prediction model with RNN-LSTM. Before modelling, we take week as time step of prediction according to the frequency of lecture release in Chinese University MOOC platform. And then, do data preprocessing and cleaning work on raw datasets provided by Chinese University MOOC, by analysing the features of learners learning behavior, fifteen features are extracted for the training and testing of model. The weekly status labels of each learner are generated with the combination of four learning status' definitions and features. The input of model consist of features and labels, 80\% of it is taken as training datasets, and 20\% of it is taken as testing datasets. The implementation of prediction model is based on Google's TensorFlow. According to the testing results, each course acquires well prediction results in the open semester.
\par
In the future, we still need to take some optimization techniques to adjust parameters of the proposed prediction model, which is helpful to improve the efficiency of training. Moreover, the prediction results of the same course in different semester still need improvement. Besides, we hope to implement transfer learning between courses.

\bibliographystyle{plain}
\bibliography{elsarticle-template}

\begin{thebibliography}{10}

\bibitem{csdnblog}
a~bird's sky.
\newblock An introduction to rnn (recurrent neural networks).
\newblock \url{http://blog.csdn.net/heyongluoyao8/article/details/48636251}.
\newblock Accessed March 12, 2017.

\bibitem{balakrishnan2013predicting}
Girish Balakrishnan and Derrick Coetzee.
\newblock Predicting student retention in massive open online courses using
  hidden markov models.
\newblock {\em Electrical Engineering and Computer Sciences University of
  California at Berkeley}, 2013.

\bibitem{dennyrnn}
Denny Britz.
\newblock Wildml, recurrent neural networks tutorial, part1-introduction to
  rnns.
\newblock
  \url{http://www.wildml.com/2015/09/recurrent-neural-networks-tutorial-part-1-introduction-to-rnns/}.
\newblock Accessed March 31, 2017.

\bibitem{gers1999learning}
Felix~A Gers, J{\"u}rgen Schmidhuber, and Fred Cummins.
\newblock Learning to forget: Continual prediction with lstm.
\newblock 1999.

\bibitem{gers2002learning}
Felix~A Gers, Nicol~N Schraudolph, and J{\"u}rgen Schmidhuber.
\newblock Learning precise timing with lstm recurrent networks.
\newblock {\em Journal of machine learning research}, 3(Aug):115--143, 2002.

\bibitem{he2015identifying}
Jiazhen He, James Bailey, Benjamin~IP Rubinstein, and Rui Zhang.
\newblock Identifying at-risk students in massive open online courses.
\newblock In {\em AAAI}, pages 1749--1755, 2015.

\bibitem{jiang2014predicting}
Suhang Jiang, Adrienne Williams, Katerina Schenke, Mark Warschauer, and Diane
  O'dowd.
\newblock Predicting mooc performance with week 1 behavior.
\newblock In {\em Educational Data Mining 2014}, 2014.

\bibitem{kay2013moocs}
Judy Kay, Peter Reimann, Elliot Diebold, and Bob Kummerfeld.
\newblock Moocs: So many learners, so much potential.
\newblock {\em IEEE Intelligent Systems}, 28(3):70--77, 2013.

\bibitem{khalil2014moocs}
Hanan Khalil and Martin Ebner.
\newblock Moocs completion rates and possible methods to improve retention-a
  literature review.
\newblock In {\em World Conference on Educational Multimedia, Hypermedia and
  Telecommunications}, volume~1, pages 1305--1313, 2014.

\bibitem{kinga2015method}
D~Kinga and J~Ba Adam.
\newblock A method for stochastic optimization.
\newblock In {\em International Conference on Learning Representations (ICLR)},
  2015.

\bibitem{kizilcec2015attrition}
Ren{\'e}~F Kizilcec and Sherif Halawa.
\newblock Attrition and achievement gaps in online learning.
\newblock In {\em Proceedings of the Second (2015) ACM Conference on Learning@
  Scale}, pages 57--66. ACM, 2015.

\bibitem{kloft2014predicting}
Marius Kloft, Felix Stiehler, Zhilin Zheng, and Niels Pinkwart.
\newblock Predicting mooc dropout over weeks using machine learning methods.
\newblock In {\em Proceedings of the EMNLP 2014 Workshop on Analysis of Large
  Scale Social Interaction in MOOCs}, pages 60--65, 2014.

\bibitem{koedinger2015learning}
Kenneth~R Koedinger, Jihee Kim, Julianna~Zhuxin Jia, Elizabeth~A McLaughlin,
  and Norman~L Bier.
\newblock Learning is not a spectator sport: Doing is better than watching for
  learning from a mooc.
\newblock In {\em Proceedings of the Second (2015) ACM Conference on Learning@
  Scale}, pages 111--120. ACM, 2015.

\bibitem{beijingNews}
Beijing News.
\newblock the pass rate of chinese university moocs is only 1.5.
\newblock
  \url{http://www.edu.cn/info/zyyyy/zxjy/mooc/201702/t20170220_1490524.shtml}.
\newblock Accessed July 28, 2017.

\bibitem{csdnblogrnnlstm}
Dark Scope.
\newblock Introduction to rnn and lstms.
\newblock \url{http://blog.csdn.net/article/details/47056361}.
\newblock Accessed July 1, 2017.

\bibitem{srivastava2014dropout}
Nitish Srivastava, Geoffrey~E Hinton, Alex Krizhevsky, Ilya Sutskever, and
  Ruslan Salakhutdinov.
\newblock Dropout: a simple way to prevent neural networks from overfitting.
\newblock {\em Journal of machine learning research}, 15(1):1929--1958, 2014.

\bibitem{taylor2014likely}
Colin Taylor, Kalyan Veeramachaneni, and Una-May O'Reilly.
\newblock Likely to stop? predicting stopout in massive open online courses.
\newblock {\em arXiv preprint arXiv:1408.3382}, 2014.

\bibitem{tian2015predicting}
Yongxue Tian and Li~Pan.
\newblock Predicting short-term traffic flow by long short-term memory
  recurrent neural network.
\newblock In {\em Smart City/SocialCom/SustainCom (SmartCity), 2015 IEEE
  International Conference on}, pages 153--158. IEEE, 2015.

\bibitem{Liuwenyan2016MOOC}
Liu Wenyan and Luo Tao.
\newblock Research on the prediction of mooc drop-out rate.
\newblock 2016.

\bibitem{whitehill2015beyond}
J~Whitehill, J~Williams, G~Lopez, C~Coleman, and J~Reich.
\newblock Beyond prediction: Toward automatic intervention to reduce mooc
  student stopout.
\newblock {\em Educational Data Mining}, 2015.

\bibitem{cnblogsadam}
xinchrome.
\newblock Adam, a random optimizer xinchrome.
\newblock \url{http://www.cnblogs.com/xinchrome/p/4964930.html}.
\newblock Accessed March 12, 2017.

\bibitem{ye2014early}
Cheng Ye and Gautam Biswas.
\newblock Early prediction of student dropout and performance in moocss using
  higher granularity temporal information.
\newblock {\em Journal of Learning Analytics}, 1(3):169--172, 2014.

\end{thebibliography}

\end{spacing}{1.0}
\end{document}